

\input harvmac

\noblackbox
\pageno=0\nopagenumbers\tolerance=10000\hfuzz=5pt
\line{\hfill CERN-TH.7422/94}
\vskip 36pt
\centerline{\bf CALCULATING $F_2^p$ AT
                SMALL $x$ AND LARGE $Q^2$}
\vskip 36pt\centerline{Richard~D.~Ball\footnote{$^\dagger$}{On leave
from a Royal Society University Research Fellowship.}
 and Stefano~Forte\footnote{$^\ddagger$}{On leave
from INFN, Sezione di Torino, Italy.}}
\vskip 12pt
\centerline{\it Theory Division, CERN,}
\centerline{\it CH-1211 Gen\`eve 23, Switzerland.}
\vskip 60pt
{\medskip\narrower
\ninepoint\baselineskip=9pt plus 2pt minus 1pt
\lineskiplimit=1pt \lineskip=2pt
\centerline{\bf Abstract}
\noindent
We show that the double asymptotic scaling of the HERA structure function
data is consistent with pre-HERA data at larger $x$, soft pomeron
behaviour at small $x$ and a sensible starting scale $Q_0$. We can thus
actually calculate $F_2^p$ at small $x$ and large $Q^2$ by evolving up
perturbatively at two loops, without any fitting.
}
\vfill
\centerline{Talk presented at ``QCD94'', Montpellier, July 1994,}
\centerline{to be published in the proceedings (Nucl. Phys. B (Proc. Suppl.))}
\vskip 20pt
\line{CERN-TH.7422/94\hfill}
\line{September 1994\hfill}

\vfill\eject
\footline={\hss\tenrm\folio\hss}


\def\frac#1#2{{{#1}\over {#2}}}

\def\smallfrac#1#2{\hbox{${{#1}\over {#2}}$}}

\def\GeV{{\rm GeV}}

\catcode`@=11 
\def\slash#1{\mathord{\mathpalette\c@ncel#1}}
 \def\c@ncel#1#2{\ooalign{$\hfil#1\mkern1mu/\hfil$\crcr$#1#2$}}
\def\lsim{\mathrel{\mathpalette\@versim<}}
\def\gsim{\mathrel{\mathpalette\@versim>}}
 \def\@versim#1#2{\lower0.2ex\vbox{\baselineskip\z@skip\lineskip\z@skip
       \lineskiplimit\z@\ialign{$\m@th#1\hfil##$\crcr#2\crcr\sim\crcr}}}
\catcode`@=12 

\def\SZP{\hbox{S0$'$}}\def\DZP{\hbox{D0$'$}}\def\DMP{\hbox{D-$'$}}
\def\szp{\hbox{(S0$'$)0}}\def\dzp{\hbox{(D0$'$)0}}\def\dmp{\hbox{(D-$'$)0}}
\def\PR{{\it Phys.~Rev.~}}

\def\PL{{\it Phys.~Lett.~}}

\def\vol#1{{\bf #1}}\def\vyp#1#2#3{\vol{#1} (#2) #3}


\nref\DGPTWZ{ A.~De~Rujula et al, \PR\vyp{D10}{1974}{1649}.}
\nref\DAS{ R.D.~Ball \& S.~Forte, \PL\vyp{B335}{1994}{77}.}
\nref\HERA{ M.~Roco \& K.~M\"uller, results from
ZEUS \& H1 presented at the 29th Rencontre de Moriond, March 1994.}
\nref\Test{ R.D.~Ball \& S.~Forte, CERN-TH.7331/94,
                {\tt hep-ph/9406385}, \PL{\bf B} (in press).}
\nref\Blois{R.D.~Ball \& S.~Forte, CERN-TH.7421/94.}
\nref\MRS{ A.D.~Martin et al, \PL\vyp{B306}{1993}{145}.}
\nref\MRSH{ A.D.~Martin et al, RAL-94-055, DTP/94/34.}
\nref\GRV{ M.~Gl\"uck et al, \PL\vyp{B306}{1993}{391}.}
\nfig\sr1{Double scaling plots of $R_F F_2^p$ vs. a) $\sigma$ and b)
$\rho$. The data are taken from ref.\HERA, and the curves are those of
the new parton distributions \dzp. The old MRS prediction \DZP\ is
shown dotted.}
\nfig\sr2{As fig.~1, but with the curves now corresponding to \dmp\
and \DMP (dotted).}

Perturbative QCD makes a definite prediction \DGPTWZ\ for the shape
of the rise of the singlet part of the structure function $F_2$ at
large $Q^2$ and small $x$, provided only that the starting distribution is
not too singular. This prediction is most striking when
expressed as a double asymptotic scaling in the two scaling variables
\DAS
\eqn\esr{
\sigma\equiv\sqrt{\ln\smallfrac{x_0}{x}\ln\smallfrac{t}{t_0}},
\qquad\rho\equiv\sqrt{\ln\smallfrac{x_0}{x}\big/\ln\smallfrac{t}{t_0}},
}
where $t\equiv\ln(Q^2/\Lambda^2)$. When $F_2$ is rescaled by a
multiplicative factor
\eqn\eR{
R_F\equiv N\sigma^{1/2}\rho e^{-2\gamma\sigma +\delta\sigma/\rho},
}
it should scale in both $\sigma$ and $\rho$ as they become large, since
the growth predicted in \DGPTWZ\ has been scaled out. The
parameter $\gamma$ which controls the rate of growth of $F_2$ is simply
related to the leading coefficient of the beta function (in fact
$\gamma\equiv\sqrt{12/\beta_0}$), while $\delta$ is an anomalous dimension.
Recent data from HERA \HERA\ are in remarkably good agreement with
this double scaling prediction, to the extent
that it may be used to directly determine $\beta_0$ by measuring the
slope of the  rise\Test. It thus now becomes imperative to
compute higher order corrections: double scaling violations.

The original calculation in \DGPTWZ\ expanded the one loop anomalous
dimensions around their leading singularity, which should dominate the
evolution at large $Q^2$ and small $x$ if there is no corresponding
singularity in the input distribution. In \DAS\ we showed that this
approximation turns the Altarelli-Parisi equation for
the gluon distribution $xg(x;t)$ into a two-dimensional wave equation with
light-cone variables $\ln(x_0/x)$ and $\ln(t/t_0)$. The (unstable)
propagation of soft boundary conditions (in particular $g(x;t_0)\sim
x^{-1}$ as $x\to 0$, the intercept of the soft pomeron being close to unity)
into the interior of the light-cone then produces the generic rise
in $F_2$. Asymptotically the details of the boundary conditions are washed
out, and only their overall normalization is left. The simplest approximation
is thus sufficient to capture the essential physics at small $x$: $F_2$ rises
because of the instabilty of gluons to gluon emission via the
triple-gluon vertex.

Of course the success of this simple picture still leaves some important
questions unanswered. In particular, is the double asymptotic scaling
behaviour consistent with the structure functions
measured (pre-HERA) at larger values of $x$? Furthermore, how good is the
leading singularity approximation to the splitting function, and how large
are the two loop corrections? In other words, it would
be useful to compute the normalization and sub-asymptotic corrections
to double scaling, not only to explain why double scaling works so well,
but also to refine the comparison with the increasingly precise
experimental data. Post-asymptotic corrections due to
singularities in yet higher orders of perturbation theory, or to higher twist
corrections from parton recombination effects, are discussed briefly in
\Blois; they will not be considered here.

To compute the normalization and sub-asymptotic corrections numerically,
we adopt the following four step algorithm:

(a) Take a set of parton distributions $\Delta$ which has been fitted
to pre-HERA data with $Q^2 \gsim 4\GeV^2$, $x\gsim10^{-2}$. For
illustration we will
use here the MRS \SZP, \DZP\ and \DMP\ distributions \MRS.

(b) Evolve $\Delta$ backwards to a starting scale $Q_0$. This scale should be
chosen sufficiently low that it is reasonable to match to soft
non-perturbative behaviour there, but not so low that perturbation theory
has become meaningless. Here we choose $Q_0 = 1 \GeV$, as in \DAS.

(c) Remove the (unphysical) small-$x$ tails of the distributions (that
is those parts with $x\lsim 10^{-2}$, and replace them the
conventional expectation
from Regge theory. In particular the glue and sea distributions should be
given soft tails, $xg(x,t_0)\sim x^{-0.08}$, to match the intercept of the
(nonperturbative) soft pomeron. This
gives new distributions, which we will call $(\Delta)0$.


(d) The new distributions $(\Delta)0$ are now evolved back up to high
$Q^2$, where they can be compared to experimental data (at low $Q^2$ such
comparison would be difficult because of contamination by higher twist
corrections). At large-$x$ they will (by construction) be
indistinguishable from the original distributions
$\Delta$. At small-$x$ they should now exhibit
double scaling, but with a definite normalization and
sub-asymptotic corrections. They can thus  be compared
directly to the HERA data.

The results of this procedure\foot{The parameters of the three
new sets of distributions functions \szp, \dzp\ and \dmp\ may be
obtained by email from {\tt rball@surya11.cern.ch}.}
are shown in Fig.~1 (the distribution \dzp;
\szp\ is very similar) and Fig.~2 (the distribution \dmp),
presented as $\sigma$ and $\rho$ scaling plots; they should be compared
with the scaling plots in \refs{\DAS,\Test}.\foot{
Note that in the $\rho$ plots most of the data points now have
$\sigma \gg 1.1$.} The original MRS
distributions \DZP\ and \DMP\ are also shown (dotted; see also \Blois).
The figures largely speak for themselves. The normalization of the
HERA data \HERA\ is correctly
reproduced, with no free parameters (unless $Q_0$ were to be considered as
such): this is not a fit to the HERA data! The subleading
and two loop corrections do not spoil double scaling: although the slope of
the $\sigma$-plots is now a little lower ($2.09\pm 0.35$ for \dzp,
$1.92\pm 0.35$ for \dmp) than at leading order (where it was $2.4$; the
data have a slope of $2.37\pm 0.16$), there is now a slight rise
in the $\rho$ plot at large $\rho$.


To show quantitatively how well the new distributions account for the data,
we also give a table of values of $\chi^2$ (for all 115 data points with
$x<0.1$). In the third column we show how the $\chi^2$ falls if the
normalization of the distributions is left free.
For comparison we give in square brackets
similar statistics for the original MRS distributions \MRS, none
of which fit the HERA data because they do not exhibit double scaling (\SZP\
and \DZP\ because $Q_0$ was too large, \DMP\ because it incorporates the
power-like singular growth inspired by the Lipatov 'hard pomeron').\foot{
The GRV distributions \GRV\ also fail; they implicitly incorparate
double scaling, but overshoot the data because their starting scale is
much too low.} The more recent H and A distributions
\MRSH, which achieve a  fit to the HERA data by introducing
two new free parameters (basically parameterizing a small admixture of
a power rise at small-$x$, thus roughly interpolating between \DZP and \DMP)
have for comparison a $\chi^2$ of $100$. From the calculations presented
here it should be clear that similar (perhaps better) results could have
been obtained with no new parameters simply by dropping $Q_0^2$ from $4\GeV^2$
to $1\GeV^2$, and there taking a soft initial distribution of the same form
as \DZP: indeed if such an approach had been originally taken in \MRS,
an extremely accurate prediction for $F_2^p$ at small $x$ would have resulted.

If using perturbation theory at scales as low as $1 \GeV$ makes
one uncomfortable, one could instead incorporate the expected asymptotic
behaviour \DGPTWZ\ by hand into a starting distribution at $4 \GeV^2$;
it really does not matter how the double scaling is produced, provided it
obtains throughout the HERA kinematic region. After all, the most
striking feature of the small-$x$ data \HERA\ is the precocious
onset of double asymptotic scaling \DAS. But if one evolves at two
loops from a soft distribution at $1 \GeV$, one can also generate
both the correct normalization and subasymptotic corrections.

\topinsert\hfil
\vbox{\tabskip=0pt \offinterlineskip
      \def\tablerule{\noalign{\hrule}}
      \halign to 350pt{\strut#&\vrule#\tabskip=1em plus2em
                   &\hfil#\hfil&\vrule#
                   &\hfil#\hfil&\vrule#
                   &\hfil#\hfil&\vrule#
                   &\hfil#\hfil&\vrule#
                   &\hfil#\hfil&\vrule#\tabskip=0pt\cr\tablerule
      &&\omit&&\omit
             &&\multispan3 \hfil Best fit Normalization \hfil
             &&\omit\hidewidth $\%$ mom.\hidewidth\cr
      &&\omit&&\omit\hidewidth $\chi^2$\hidewidth
             &&\omit\hidewidth $\chi^2$\hidewidth
             &&\omit\hidewidth Rel. Norm.\hidewidth
             &&\omit\hidewidth in glue \hidewidth\cr\tablerule
&&DAS\DAS  &&               && $107$~~~~~~   &&
                         && $ 33\%$~~~~~~ &\cr\tablerule
&&\szp     && $115$~$[309]$ && $105$~$[230]$ && $1.04$~$[1.13]$
                         && $40\%$~$[44\%]$ &\cr
&&\dzp     && $126$~$[263]$ && $106$~$[213]$ && $1.06$~$[1.10]$
                         && $41\%$~$[44\%]$ &\cr
&&\dmp     && $114$~$[323]$ && $112$~$[185]$ && $0.98$~$[0.87]$
                         && $38\%$~$[43\%]$ &\cr\tablerule}}
\hfil\bigskip
\centerline{Table.}
\bigskip
\endinsert

{
\smallskip\noindent
Acknowledgements: We are particularly grateful to R.K.~Ellis for a copy
of his efficient two-loop evolution code. We would also like to thank
S.~Catani, F.~Hautmann, Z.~Kunszt, J.~Kwiecinski, P.V.~Landshoff,
R.G.~Roberts and D.A.~Ross for discussions.}

\bigskip
\listrefs
\listfigs
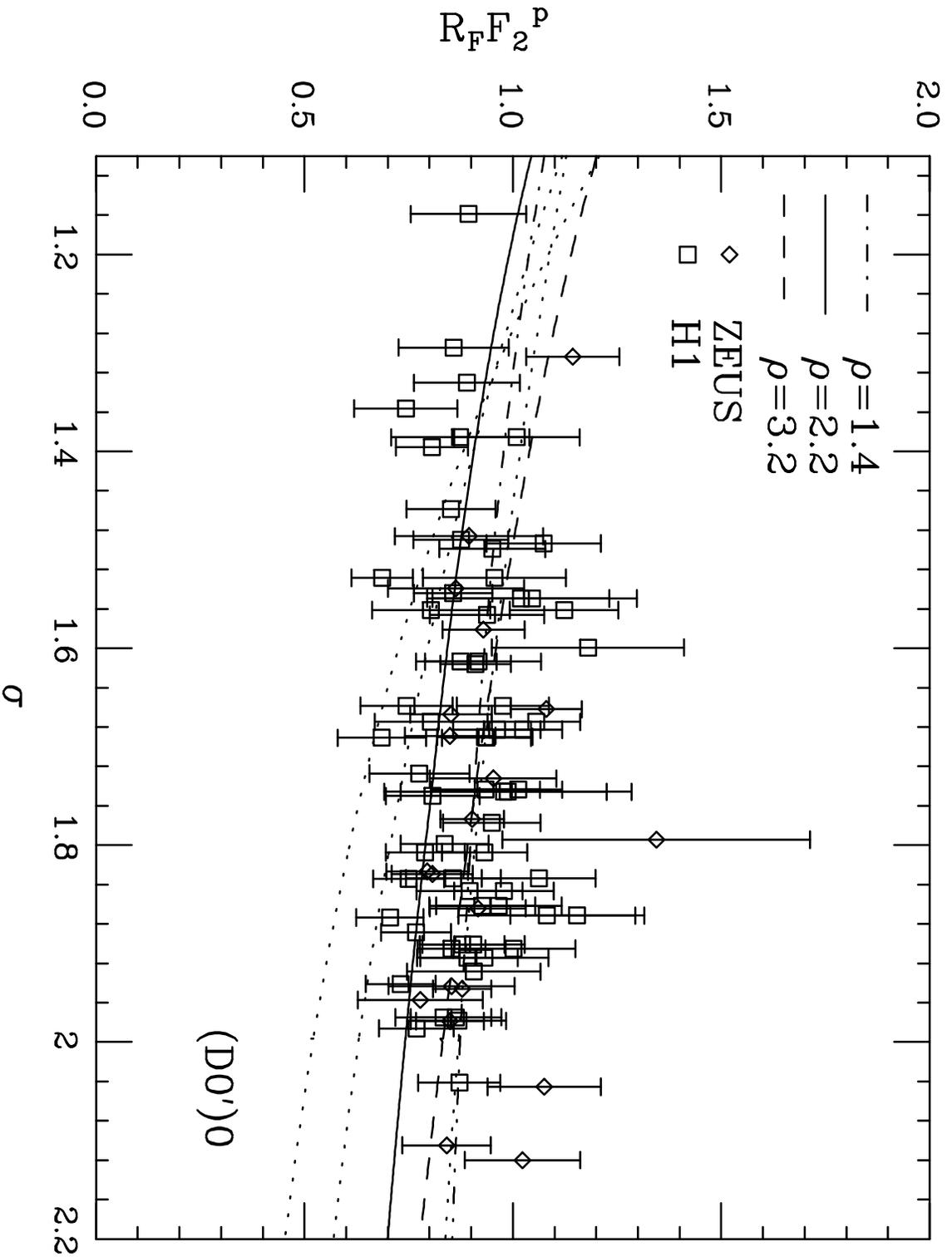
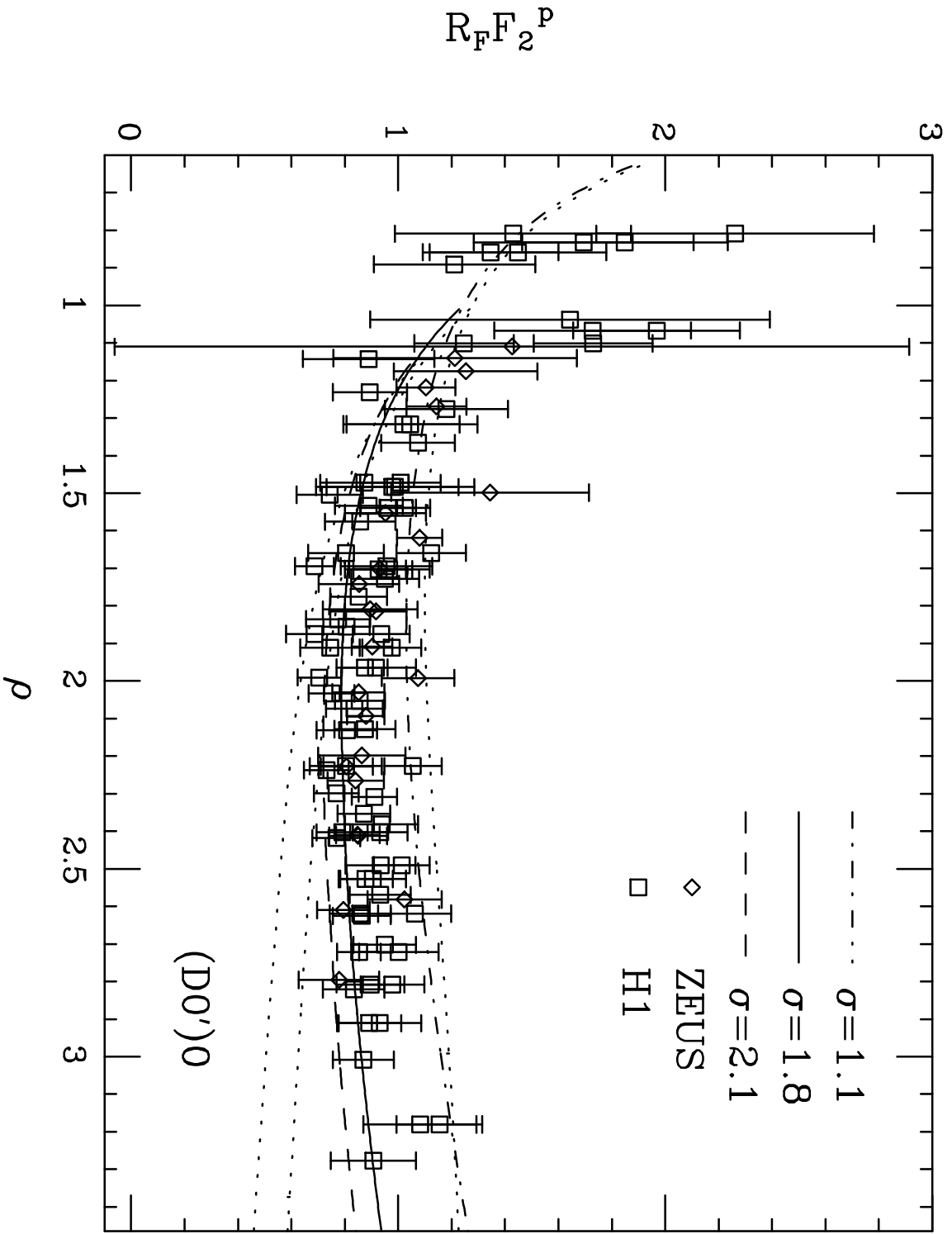
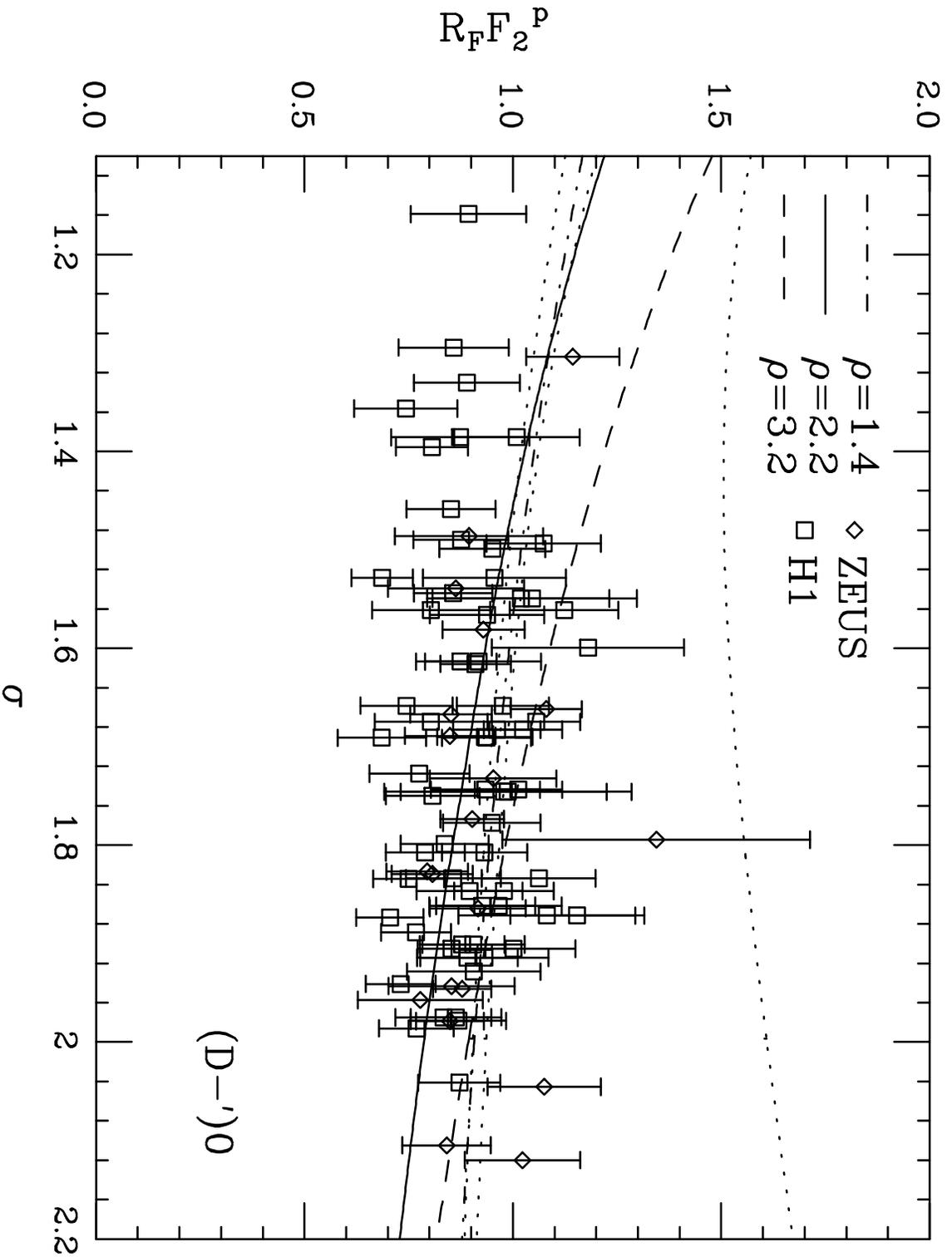
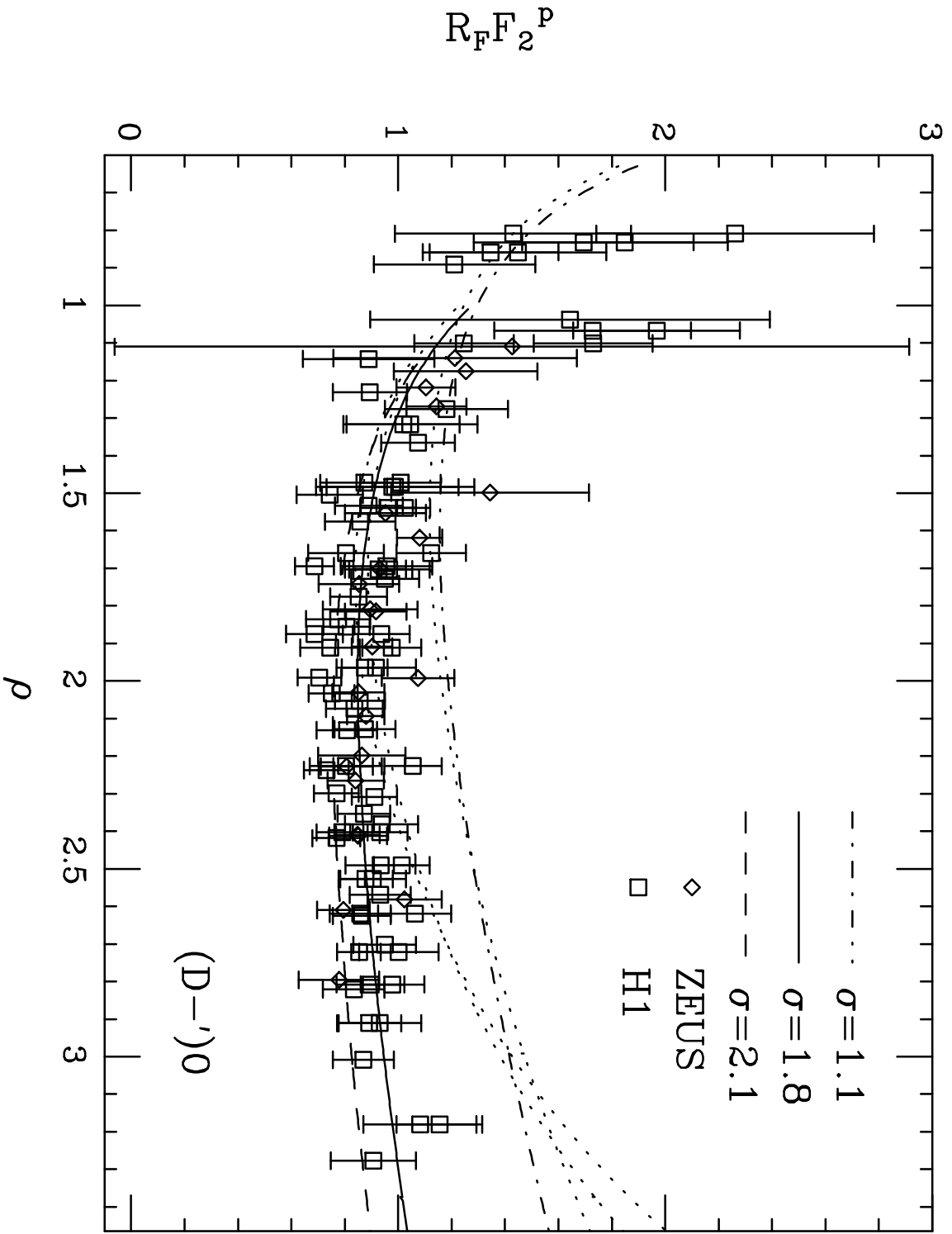
\end